\documentclass[preprint,12pt]{elsarticle}

\usepackage{amssymb}
\usepackage{amssymb}
\usepackage{lipsum}
\usepackage{graphicx}
\usepackage{booktabs}
\usepackage{amsthm}

\usepackage{lineno}

\begin{document}

\begin{frontmatter}



\title{From the revelation of the nature of inertial forces to questioning the proposition of dark energy}


\author[aff1]{ChiYi Chen}

\affiliation[aff1]{organization={School of Physics},
            addressline={Hangzhou Normal University},
            city={Hangzhou},
            postcode={310036},
            state={Zhejiang},
            country={China}}



\begin{abstract}
Starting from the revelation of the nature of inertial forces, this article discusses the subdivision of the basic physical concept of space-time and raises questions about the metric of standard cosmology.  A new form of particle dynamics equation which is strictly applicable to any translational reference frame is rigorously derived within the framework of Newtonian dynamics. This necessary brings three important implications. Firstly, a comprehensive and detailed derivation of the generalized dynamics equation strongly suggests that the fundamental concept of space-time in physics should distinguish between the absolute background of space-time and the relative inherent units of space-time. Secondly, the inertial force is successfully revealed as the real force acting on the reference object, which directly conflicts with Einstein's strong equivalence principle. Thirdly, Einstein's idea in general relativity to eliminate the dependence on inertial reference frames loses logical support. Based on these new physics and results, a properly adjusted physical picture of space-time is introduced. The coordinates in the reference frame as the background on which the gravitation will be geometrized have clear physical meanings, and they should actually be defined according to the running rate of the observer's own clock and the background length of the observer's own ruler at the present moment, and then duplicated into full space-time. Consistently, the energy-momentum of the matter should be causally and symmetrically accounted for relative to the reference origin. After that, the curvature of space-time reflects the difference between the local clocks (or rulers) in the gravitational field and the observer's clocks (or rulers) in the background reference system. In this spirit, the gravitational redshift effect in the Schwarzschild metric is also successfully re-interpreted. Correspondingly, the FRWL standard cosmological metric is necessary modified. A new factor $b(t)$ must be introduced into the time term $dt$ to represent the time dilation effect caused by the increase in the potential energy of cosmic average gravitation on the longitudinal direction of time. Qualitative analysis indicates that the proposition of dark energy can be alleviated, and at the very least, a re-analysis should be conducted based on the proposed new cosmological metric.
\end{abstract}




\end{frontmatter}


\section{Introduction}
\label{intro}
Cosmology has gradually evolved into a precise science\cite{bondi,kolb}. However, unlike other branches of physics, cosmology must pay more attention to the correctness of the entire theoretical framework, especially the accuracy of its basic observational theory, as it is not always possible to design experiments for verification\cite{WMAP,perlmutter}.

The observations of the universe and the theoretical modeling necessitate the establishment of the standard cosmology based on the cosmological principle, which asserts the homogeneity and isotropy of the universe on large scales. It is important to highlight that even the observational theory of cosmology is also established on this cosmological principle\cite{Riess,perlmuer}. The FRWL cosmological metric, as the fundamental basis of cosmological observation theory, is precisely constructed based on this principle\cite{Weinberg,Ohanian},
\begin{equation}
ds^2=-dt^2+a^2(t)[\frac{dr^{2}}{1-kr^{2}}+r^{2}d\theta^{2}+r^{2}sin^{2}\theta d\phi^{2}].
\end{equation}
But after years of exploring dark energy models, when we redirect our focus to the foundation of the dark energy proposition\cite{Turner}, we realize that the FRWL metric is not actually the most general expression of the cosmological principle. Firstly, from a mathematical standpoint, the most complete metric has the following form,
\begin{equation}
ds^2=-b^{2}(t)dt^2+a^2(t)[\frac{dr^{2}}{1-kr^{2}}+r^{2}d\theta^{2}+r^{2}sin^{2}\theta d\phi^{2}].
\end{equation}
Secondly, from a physical standpoint, according to the solar gravitational experiments, light signals transmitted from the surface of the sun to an observer on Earth experience an increase in gravitational potential energy, resulting in a redshift in the received light signals\cite{Weinberg,Ohanian}. Based closely on this physics, we know that matter density in the universe has changed significantly from the beginning to the present day, so the average gravitational potential energy of the universe, even though it is uniform in space, varies over time. Let us conduct a thought experiment. In a purely gravitational universe, the expansion process should slow down, with the decrease in its kinetic energy being transformed into an increase in the average gravitational potential energy of the universe! According to the gravitational time dilation effect already confirmed by solar gravitational experiments\cite{tower,aircraft,Bailey,rocket}, there should be variations in the running rate of the local clock at any commoving point in the universe at different moments along the time axis. Therefore, considering the long cosmic evolutionary history studied in cosmology, the time coordinates in the cosmological metric must strictly differentiate the running rate between the Earth observer's clock at the present moment and the commoving local clock at an early moment. Therefore, a complete and self-consistent cosmological metric should be formulated by the above equation (2). Now the factor $b(t)$ represents the change in the running rate of the local clock in the longitudinal direction of time, similar to the gravitational redshift effect in the solar gravitational experiments.

Why is there no $b(t)$ factor in the standard FRWL cosmological metric? In traditional textbooks of cosmology, it is generally considered that due to the principle of general covariance, the $b(t)$ factor is absorbed as $b(t)dt=d\tau$\cite{Weinberg}. So, does this imply that our prior understanding of physics was incorrect, or that the principle of general covariance is flawed?  The pursuit of truth has led to an investigation of the physics of gravitational redshift effect in cosmology, based on a strong belief that physics cannot simply disappear. This pursuit has involved both top-down phenomenological analysis and bottom-up fundamental exploration\cite{Chenc,Chench,Chenchi}.

On the path of bottom-up fundamental exploration, we questioned the principle of general covariance, and thereby the Einstein strong equivalence principle, finally leading to a questioning of the nature of inertial forces\cite{Einstein,Hanoch,equivalence,Liu}. As a physical concept, the inertial force first originated from the application of Newton's second law in non-inertial reference frames\cite{Isaac,Cheng,Zhou}. Therefore, within the framework of Newtonian dynamics, the possibility of further improvement on Newton's second law has been a subject of my personal long-term contemplation\cite{Weinberg,Liang,Gaoxie,Saunders}. Coincidentally, after six years' careful exploration, during a visit to the KITP in Beijing, it was discovered that Newton's second law can indeed be enhanced within the framework of classical mechanics\cite{Chenc,Chench},
\begin{equation}
{\textbf {\emph F}}\vert_{p}-\frac{m_{p}}{m_{o}}{\textbf {\emph F}}\vert_{o}=m_{p}{\textbf {\emph a}}\vert_{p-O}.
\end{equation}
Here, $o$(in lowercase) in principle can be an arbitrary reference object that naturally defines the reference origin of a translational reference frame $O$(in uppercase). The adjective word "translational" just means that the coordinate axis of $O$ does not rotate relative to the background of cosmic space but its reference origin can be in arbitrary acceleration including the circling motion. Therefore, the nature of the inertial force in translational reference frames is revealed as the following expression\cite{Chench},
\begin{equation}
{\textbf {\emph f}}\vert_{inertial}=-\frac{m_{p}}{m_{o}}{\textbf {\emph F}}\vert_{o}.
\end{equation}
This expression is confirming our initial suspicion that the inertial force is not equal to the gravitational force, and Einstein's strong equivalence principle does not hold.

It is precisely because Einstein's strong equivalence principle is no longer valid. As a result, the local clocks in the local inertial reference frames will no longer be running-rate invariant but will change under the influence of the gravitational field. This means that the physical picture of gravitation being geometrized should be changed. In particular, the observational theory in general relativity must be modified. Given that the co-moving points in cosmology are equivalent to the local inertial reference frames in a free-falling gravitational field, it can be demonstrated that the running rate of clocks at the co-moving coordinates will be changeable during the long-time evolution of the universe. This, in turn, confirms the necessity of introducing the factor $b(t)$ in the cosmological metric, as shown in the equation (2).

On the path of top-down phenomenological analysis, we also investigated the impact of gravitational time dilation on the dark energy proposition\cite{Chenchi}. Once again, coincidentally, we discovered that the gravitational time dilation effect, analogous to that in the Schwarzschild metric, can at least partially solve the so-called dark energy puzzle, and may even have the potential to refute the dark energy proposition.

Therefore, the structure of this article is as follows: the first part serves as the introduction, the second part presents the logical necessity of modifying Newton's second law within the framework of classical mechanics. It is emphasized that Newton's second law actually implies a causal asymmetry in its formal logic. The recognition of this error is a crucial step in the discovery of the generalized dynamics equation. The third part presents the inevitable correctness of the generalized dynamics equation within the framework of classical mechanics. It is surprising that the generalized dynamics equation can be derived rigorously by invoking Newton's second law in an inertial reference frame without any additional assumptions. But the deduction result formally avoids a direct dependence on the inertial reference frame. Based on the new physics in Sections 2 and 3, we will obtain three new results. In the fourth part, in order to completely eliminate the reliance on the concept of inertial reference frames, the generalized dynamics equation is re-derived based on the spirit of causal symmetry. The only one logical foundation that needs to be borrowed in the derivation process is that any particle must occupy an objective position in the background of cosmic space at any moment. The objectivity of this position strongly implies the absoluteness of the background of cosmic space, therefore introducing a deeper distinguishment in basic physical concepts of space-time. The fifth part highlights that the nature of inertial forces has been thoroughly revealed as the real force acting on the reference object, and this physical interpretation directly conflicts with Einstein's strong equivalence principle. The sixth part points out that the fundamental physics of generalizing the principle of relativity involves maintaining causal symmetry on both sides of new dynamics equations. Mathematically, the construction of a general principle of relativity requires the introduction of forces acting on four non-coplanar reference particles that determine the degrees of freedom of any arbitrary moving reference frame. Due to the extreme complexity of the mathematics involved, the question of whether a general principle of relativity can be realized in particle dynamics is still under pending. Following the new results in Sections 4, 5 and 6, this article makes two proposals. The seventh part discusses a properly modified physical picture of space-time when gravitation is geometrized and then examines its ability to explain the gravitational redshift effect in solar gravity experiments. The eighth part, based on the adjusted space-time physical picture, naturally provides the most general cosmological metric under the cosmological principle and analyses its physical implications for cosmological dynamics. The ninth part presents a conclusion.

\section{New Physics 1/2: The logical necessity of modifying Newton's second law within the framework of classical mechanics}
The predictability of physical laws is their most vital characteristic. In practical experiments, cause and effect should be separable, and the causal correspondence between them is a fundamental requirement of physical laws\cite{Kinsler}.

In Newtonian mechanics, the law of causality is used as the basis to determine the calculation formula of common forces. It is not Newton's second law, which requires the exact inertial reference frame to be found and the total net force to be fully counted. The law of causality should be summarized directly from a large number of classical mechanics experiments\cite{Isaac}. For example, referring to a classical mechanics experiment, we typically analyse the dynamics of a moving object in relation to its previous mechanical state\cite{Ferrel}, with the ground (or laboratory) reference frame being selected. This involves examining a causal difference equation between the previous and post states, which is determined by the newly imposed force and the resulting relative acceleration\cite{Hoek}. In this sense, what law of causality is relied upon in practice is just an empirical law, which can be described by the following difference equation:
\begin{equation}
\Delta{\textbf {\emph F}}\cong m\Delta{\textbf {\emph a}}.
\end{equation}
Here $\Delta{\textbf {\emph F}}$ represents the new imposed force compared with the previous state, and $\Delta{\textbf {\emph a}}$ represents the resulting acceleration increment compared with the previous state. Essentially, this difference equation (the differential causality may be expressed as $d{\textbf {\emph F}}\cong md{\textbf {\emph a}}$) has historically been used as the basis for studying the calculation formula for common forces, such as gravitational force, friction, and elasticity\cite{Krantz}. Once the formula for calculating forces has been summarised from specific experiments, the equation of dynamics can be tested in other general cases. Therefore, the particle dynamics is essentially a causal law about force and acceleration. Force should be the cause, and acceleration should be the effect\cite{Kinsler}.

Furthermore, it is necessary to convert the difference equation($\Delta{\textbf {\emph F}}=m\Delta{\textbf {\emph a}}$), which serves as an empirical law, into a theoretical formula. The force term in the theoretical formula must account for all the forces acting on the particle. This is important because if we fail to do so, we will not know which forces should be taken into account when applying the formula in a new situation.

Starting from the most basic logic, let's see what problems arise if we directly apply Newton's second law to non-inertial reference frames($O$). In this case the theoretical formula of Newton's second law can be tentatively denoted as ${\textbf {\emph F}}\vert_{p}?\neq m_{p}{\textbf {\emph a}}\vert_{p-O}$. The force is the cause and the acceleration is the effect. The cause(${\textbf {\emph F}}\vert_{p}$) depends only on the object under study($p$), while the effect(${\textbf {\emph a}}\vert_{p-O}$) depends on both the object under study($p$) and the reference frame($O$). But for any actual frame of reference, its physics must inevitably be traced back to the reference object(s) defining it. A translational reference frame, for example $O$, can be defined with only one reference object, which can be denoted as $o$
(Here a reference object and its corresponding reference frame are represented by lower-case and upper-case letters to distinguish and link them respectively). Now we can analyse the formal logic of Newton's second law from a different perspective, omitting the details of force and acceleration, and focusing solely on the variables representing the particles. There is only one formal variable of cause(${\textbf {\emph F}}\vert_{p}$), which is $p$, and two formal variables of effect(${\textbf {\emph a}}\vert_{p-O}$), which are $p$ and $o$. However, the selection of the reference object $o$ is completely independent of $p$. Therefore, based on formal logic,
\begin{eqnarray}
{\textbf {\emph F}}\vert_{p} &\Longleftrightarrow\qquad \qquad  Cause(p)&:p\qquad  \nonumber  \\
{\textbf {\emph a}}\vert_{p-O}&\Longleftrightarrow Effect(p,O)\Longleftrightarrow Effect(p,o)&:p, o
\end{eqnarray}
It is evident that the cause and effect described in Newton's second law do not satisfy a one-to-one correspondence (as the formal variables are different). This is referred to as causal asymmetry in this article.

In practical reference frames, it is necessary to consider not only the forces acting on the object under study but also the forces acting on the reference object. This is because any physical reference frame is defined according to the real reference object. It is important to note that causal correspondence is a fundamental requirement that applies to all forms of particle dynamics, including classical mechanics and relativistic mechanics. This requirement must be satisfied in general formulism.

\section{New Physics 2/2: The inevitable correctness of the generalized dynamics equation within the framework of classical mechanics}
As a physical reference frame must be located within the real universe, in some cases there also arises a question as to whether the reference object can be identified\cite{Einstein,Zhou}. When using the ground(or laboratory) reference frame, in fact, any stationary object in the ground(or laboratory) can be selected as the reference object. Acceleration is measured with respect to any object at rest on the ground or in the laboratory. The range of the reference object can be arbitrarily determined as long as it can be regarded as a particle. The properties of the reference object, such as its mass, do not affect the applicability. However, a real object must exist at that point in order to validate physical measurements. Even for the center-of-mass reference frame, it is equivalent to selecting all the objects in the whole system as reference objects, only with a mathematical treatment for determining the spatial location of the representative reference ``object"(i.e., the definition for the coordinates of the center-of-mass). When studying the motion of the center-of-mass relative to the external environment, it is necessary to consider the total of external forces acting on all the particles in the system. That is to say, to calculate the force 'acting' on the center-of-mass, the center-of-mass reference frame must be boiled down to substantive reference objects.

Once the reference object has been identified, the motion of the reference frame must be attributed directly to that of the reference object, and the motion of the reference object must be attributed directly to the force acting on it. Besides, according to natural philosophy, what is chosen as the object under study and what is chosen as the reference object is essentially an artificial choice made by our human consciousness, and at the level of the fundamental law of physics, the object under study and the reference object are on a completely equal status. Based on this profound understanding, we present a logical derivation of the generalized dynamics equation (3) by invoking the concept of inertial reference frames, to which Newton's second law strictly applies\cite{Chenc,Chench}. The great way is simple as follows. In theory there exists an inertial frame of reference($\Sigma$), both an arbitrary object under study($p$) and an arbitrary reference object($o$) should in principle obey the same Newton's second law:
\begin{eqnarray}
{\textbf {\emph F}}\vert_{p}=m_{p}{\textbf {\emph a}}\vert_{p-\Sigma},  \\
{\textbf {\emph F}}\vert_{o}=m_{o}{\textbf {\emph a}}\vert_{o-\Sigma}.
\end{eqnarray}
It can be inferred that ${\textbf {\emph F}}\vert_{p}$ and ${\textbf {\emph F}}\vert_{o}$ respectively denote the total forces acting on the object under study $p$ and the reference object $o$. Both of them can in principle be calculated directly from the known theoretical formulas of common forces. We introduce the corresponding translational reference frame $O$ by taking the reference object $o$ as the reference origin even $o$ is any a real object. Then the kinematical transformation between inertial and translational reference frames is invoked:
\begin{equation}
{\textbf {\emph a}}\vert_{p-\Sigma}-{\textbf {\emph a}}\vert_{o-\Sigma}=({\textbf {\emph a}}\vert_{p-o})_{\Sigma}={\textbf {\emph a}}\vert_{p-O}.
\end{equation}
The last step in the above equation just requires the corresponding reference frame $O$ must be translational, i.e., there is no rotation of the coordinate axes with respect to the inertial reference frame $\Sigma$. Where $({\textbf {\emph a}}\vert_{p-o})_{\Sigma}$ is the relative acceleration between $p$ and $o$(by default in the inertial reference frame $\Sigma$), which is equivalent to the acceleration in the translational reference frame $O$ with the reference origin $o$: ${\textbf {\emph a}}\vert_{p-O}$. After that, both sides of above equations (7) and (8) are divided by their masses respectively and subtracted from each other. As a result, within the original framework of Newtonian dynamics, without the need for any additional assumptions, the new formulation of particle dynamics equation (3) that is directly applicable in the translational frames of reference is necessarily derived\cite{Chenc,Chench},
\begin{equation}
\frac{{\textbf {\emph F}}\vert_{p}}{m_{p}}-\frac{{\textbf {\emph F}}\vert_{o}}{m_{o}}={\textbf {\emph a}}\vert_{p-O}.
\end{equation}
It is clear that the roles of $p$ and $o$ in the above equation are symmetrical and can be interchanged while keeping the equation (10) invariant. This perfectly confirms the natural philosophical principle as pointed out earlier: $p$ and $o$ have equal status in the formulism of the fundamental law of dynamics. For the sake of distinction, the equation (10) is called as the generalized dynamics equation. This is the correct expression that Newtonian dynamics should have.

In the light of the above derivation, it might be questioned that in the special theory of relativity, the theoretical dynamics equation is ${\textbf {\emph f}}=d{\textbf {\emph p}}/dt$. However, the reformulation presented in this derivation begins with ${\textbf {\emph F}}\vert_{p}=m_{p}{\textbf {\emph a}}\vert_{p-\Sigma}$. This is reasonable because, in the context of classical Newtonian mechanics, the former formula (${\textbf {\emph f}}=d{\textbf {\emph p}}/dt$) is only more applicable to the case of variable mass than the latter (${\textbf {\emph F}}\vert_{p}=m_{p}{\textbf {\emph a}}\vert_{p-\Sigma}$). In the classical low-speed case, such as the rocket launching problem, the variable mass problem arises, but it can be categorized the separation and relative motion between particles in a system of particles\cite{Zhou}, rather than the dynamics of a single particle. Thus, the fundamental equation for particle dynamics in classical Newtonian mechanics is still ${\textbf {\emph F}}\vert_{p}=m_{p}{\textbf {\emph a}}\vert_{p-\Sigma}$, and ${\textbf {\emph f}}\vert_{p}={d\textbf {\emph p}}\vert_{p-\Sigma}/dt $ can be regarded as an effective form introduced when the single particle dynamics is extended to the system of particles. In contrast, the relativistic form of particle dynamics, ${\textbf {\emph f}}\vert_{p}= {d\textbf {\emph p}}\vert_{p-\Sigma}/dt $, is preferred in the special theory of relativity. This is because the mass of a particle can change with speed, and its physical origin can be attributed to the principle of the constant speed of light. However, even in relativistic mechanics, the most fundamental starting point is $ {F}_{\mu}= m_{0}d^{2}{x}_{\mu}/d\tau^{2}$. The relativistic form ${\textbf {\emph f}}\vert_{p}= {d\textbf {\emph p}}\vert_{p-\Sigma}/dt $ is derived from this starting point by applying it to a specific reference frame. If the principle of the constant speed of light can further be confirmed to hold in an arbitrary translational reference frame, the existing relativistic dynamics should also be transformed to meet the requirement of causal correspondence without affecting their successful applications.

\section{New Results 1/3: The logical reconstruction of generalized dynamics equation strongly implies a fundamental division between the background of space-time and the inherent units of space-time}
However, in Section 3, although the final result shows that the generalized dynamics equation no longer directly depends on the concept of inertial reference frames, it is evident that this concept was still used in the derivation process. Therefore, the physical logic of the generalized dynamics equation should be reorganized to achieve a naturally consistent understanding.

Let's bring this issue back to the starting point, how to establish the correct theoretical formula based on the causality given by the empirical law($\Delta{\textbf {\emph F}}\cong m\Delta{\textbf {\emph a}}$)?  The key is to ensure that the force and acceleration of the reference object satisfy the causal symmetry on both sides of the new formula of particle dynamics. Now, from the point of view of causal correspondence, we must think about what the acceleration due to the total force acting on a particle does mean. At any given moment, the total force acting on a particle under study should be objective and not change with the reference frame. Therefore, the corresponding result must also be objective, that is, independent of the choice of reference frames.  A complete objective acceleration of the particle under study can only be expressed as the acceleration in the background of cosmic space,
\begin{equation}
{\textbf {\emph F}}\vert_{p}=m_{p}\frac{d^{2}}{dt^{2}}\Omega\vert_{p}.
\end{equation}
The background of cosmic space here refers to the void that remains in cosmic space after the removal of all evolvable things. The symbol $\Omega$ is used here specially to indicate the objective position of the particle in the background of cosmic space. The objectivity of the position here simply means that the position has nothing to do with any artificial choice made by the human mind. It is worth pointing out that, up to now, the frame of reference in the full sense has not been artificially introduced, so the basic units of space-time used in the equation (11) should rather be understood as the background units which is mathematically introduced. The standard of the background units is yet to be defined by the observer according to his inherent physical phenomena. However, regardless of the standard in question, it is uniform everywhere. In other words, the specific value of each term in the equation (11) is subject to the future definition of the inherent units of space-time, but does not affect the validity of the equation (11). In physics, the objective position is exactly analogous to the concept of ``event" in the special theory of relativity before specific coordinate values have been assigned to it\cite{Einstein}. The objective position of every particle at every moment actually constitutes an event. Similarly, in the special theory of relativity, every event itself is assumed to have an objective position in the background of space-time so that the coordinate values of the same event in different inertial reference frames can be related  by the famous Lorentz transformation.

Digging deeper, the fact that a particle or event has an objective position in the background of cosmic space must mean that the background of cosmic space is absolute\cite{Ohanian,Kinsler}. If the background of cosmic space is not absolute, we can't even determine the rotation of any reference frame, and we don't know how to calculate the speed of rotation of coordinate base vectors: $d\textbf {\emph e}_{i}\vert_{O}/dt$. In order to be compatible with the experiments of relativistic physics, it is necessary to minimize the extent to which the concept of absoluteness holds. It is proposed to divide the concept of space-time into two levels: the one is the background of space-time, and the other is the inherent units of space-time. The inherent units of space-time is actually defined by the observer according to the physical phenomena inherent in the natural material world, so the inherent units of space-time is naturally susceptible to certain kinds of interactions and could be relative\cite{BIPM}. In contrast, in mathematics, the background of space-time, which reflects the length of the space-time inherent units, must be absolute. In physics, the background of space-time itself is not a specific matter or energy, there is no interaction acting on it. Therefore, it is really natural for the background of space-time to be absolute\cite{Machprinciple}. As the dialectical relationship of the unity of opposites, the inherent units of space-time can naturally be understood as the length of the line segment intercepted on the absolute background of space-time by the physical phenomena that define the inherent units of space-time. In this sense, the most natural and simple assumption about the absoluteness is simply that only the background of cosmic space is absolute.

Although any a particle has an objective position in the background of cosmic space, the objective position itself cannot be measured directly.  What we can actually measure is the difference between two objective positions. Therefore, we have to introduce at least one reference object(denoted by $o$), which in principle can be chosen arbitrarily from the real universe. Now we introduce the most basic way to naturally construct a space vector in the background of cosmic space,
\begin{equation}
({\textbf {\emph r}}\vert_{p-o})_{background}=\Omega\vert_{p}-\Omega\vert_{o}.
\end{equation}
Following this, a particle dynamics equation can be constructed for direct use by observers. It is important to note that all objects in the universe should be subject to the same basic law of dynamics. So for the actual reference object $o$, its dynamics also satisfies,
\begin{equation}
{\textbf {\emph F}}\vert_{o}=m_{o}\frac{d^{2}}{dt^{2}}\Omega\vert_{o}.
\end{equation}
The reference object $o$ is defined as the reference origin of a non-rotating reference frame $O$ (i.e. translational). The selection of a reference frame is intended to enable a relative measurement of acceleration, and as a result, forces should also be counted relatively\cite{Chenc,Chench},
\begin{eqnarray}
\frac{{\textbf {\emph F}}\vert_{p}}{m_{p}}-\frac{{\textbf {\emph F}}\vert_{o}}{m_{o}}&=& \frac{d^{2}}{dt^{2}}[\Omega\vert_{p}-\Omega\vert_{o}]=\frac{d^{2}({\textbf {\emph r}}\vert_{p-o})_{background}}{dt^{2}}  \cr
&=& ({\textbf {\emph a}}\vert_{p-o})_{background}={\textbf {\emph a}}\vert_{p-O}.
\end{eqnarray}
The generalized dynamics equation (10) is therefore proven again through a self-consistent derivation, as shown by equations from (11) to (14). The equation (10) is applicable to any translational reference frame $O$. Newton's second law is just an extreme case of the generalized dynamics equation (10) when the total (net) force acting on the reference object is identically equal to zero(${\textbf {\emph F}}\vert_{o}\equiv 0$). Therefore, in essence, the generalized dynamics equation and Newton's second law are not equivalent\cite{Chenc,Chench}. The generalized dynamics equation complements an independent term that is omitted by Newton's second law, in the "definite integral" process from their common basis---the differential causality( $d{\textbf {\emph F}}\cong md{\textbf {\emph a}}$).


\section{New Results 2/3: The explanation of the nature of inertial forces completely contradicts Einstein's strong equivalence principle}
From a physical perspective, the basis of Einstein's strong equivalence principle is investigated. Compared with Newton's second law (7), the second term of the generalized dynamics equation (10) just naturally explains the extra term introduced by the transformation of reference frames for Newton's second law --- the inertial force. The physical nature of the inertial force is revealed as\cite{Chench},
\begin{equation}
{\textbf {\emph f}}\vert_{inertial}=-\frac{m_{p}}{m_{o}}{\textbf {\emph F}}\vert_{o}.
\end{equation}
It can be seen that the nature of inertial force is the mass-ratio weighted real forces acting on the reference object, which can be the gravitational force or other common forces. It is important to note that these forces do not act on the object under study at all, but on the reference object. As the concept of the inertial force is originally rooted in the traditional framework of Newtonian dynamics\cite{Liu,Gaoxie}, finding an exactly equivalent correspondence or physical substitution under the same framework, is actually the most thorough way to fundamentally solve the inertial force puzzle. This goal is achieved with the explicit explanation provided in (15). It is proven that the nature of inertial forces is not physically equivalent to the gravitational force. Einstein's strong equivalence principle is therefore provided solid evidence to the contrary\cite{Hanoch}. As a result, Einstein's approach to eliminating the direct dependence on inertial reference frames in his general theory of relativity lacks logical support\cite{Weinberg}.

\section{New Results 3/3: The causal symmetry required in the construction of the principle of relativity suggests abandoning the general principle of relativity}
From a mathematical perspective, the ability of particle dynamics to remain the formula invariant under the transformation between arbitrary reference frames can be qualitatively analyzed based on the requirement of causal correspondence, given the successful reformulation of particle dynamics in translational reference frames. It is commonly understood that a physical reference frame that is arbitrarily moving (including translating and rotating) is typically attached to a rigid body that is also arbitrarily moving. To determine all degrees of freedom in the kinematics of this arbitrarily moving reference frame, at least four non-coplanar reference particles fixed on this rigid body must be selected and taken into account. Therefore, when constructing the most general formula for dynamics, it is necessary to incorporate all the forces acting on these four non-coplanar reference particles to maintain the causal symmetry on both sides. However, achieving a unified formula under current mathematical knowledge is challenging\cite{Chench}. Further research is required in the future. From a metaphysical perspective, in order to investigate the dynamics of a single particle in an arbitrarily moving reference frame, it is necessary to introduce four non-coplanar reference particles' forces must be introduced simultaneously into the formula of dynamics. This is also not cost-effective. Therefore, it is suggested that the principle of relativity for particle dynamics should be generalized at most to all translational reference frames. For the rotational part, the rotating reference frame can be used in principle. However, it is first necessary to perform a coordinate transformation into the adjoint translational reference frame. Following this transformation, the generalized dynamics equation (10) can be applied.

Therefore, through a comprehensive analysis of both physics and mathematics in Sections 5 and 6, we conclude that the general principle of relativity, at least at the level of particle dynamics, should be temporarily abandoned. Furthermore, in conjunction with a more precise division of the physical concept of space-time, namely that the absolute background and the relative inherent units serve as a unity of opposites, we introduce two proposals.

\section{New Proposal 1/2: Necessary adjustments to the physical picture of space-time and its gravitational geometrization}
On one hand, the absolute background of cosmic space does not conflict with the physical logic in the special theory of relativity. Firstly, the Lorentz transformation in the special theory of relativity provides a numerical relationship between the space-time intervals of the same two events observed in different inertial reference frames. However, the lengths of all intervals must be understood based on a background to make a meaningful comparison possible. Secondly, only if an event has an objective position in the background of space and time, then the coordinate values of the event in different inertial reference frames can be linked by the Lorentz transformation.
In the physical picture of the general theory of relativity, it has been verified that the gravitational force has a time dilation effect. That is, the length of inherent units of local clocks, compared with that of the observer's clock, changes due to the variation of gravitational fields to which local clocks are subjected. But on deeper reflection, how can such a change in the length of inherent units of the local clock (or ruler) be manifested? Only if there is an absolute background, then the change in length can be manifested and therefore has an objective meaning. In practical terms, the length of the inherent unit of a local ruler is understood as the spatial line segment intercepted from the background of cosmic space since the specific physical phenomena occurring on this local ruler defines the inherent unit. Therefore, local rulers will not have the same length of inherent units if they are under different gravitational fields\cite{Chenchi}.

On the other hand, as argued in the Sections 5 and 6, it is necessary to withdraw Einstein's strong equivalence principle and general principle of relativity. Therefore in a more conservative geometrical theory of gravitation, the physical logic for eliminating the direct dependence on inertial reference frames must be modified and adapted accordingly. In order to capture the physical essence of gravitation being geometrized, the most basic physical logic is listed below,\\
1, The concept of space-time should be divided into two levels, namely the inherent units of space-time and the background of space-time. The inherent units of space-time are defined by periodic physical phenomena of specific matter and are therefore subject to actual interactions, making them relative. In contrast, the background of space-time is not constituted by any form of matter and therefore remains unaffected by any interactions, making it absolute.\\
2, The correct geometrization of gravitation should be based on the inherent units of space-time defined by the observer's clock and ruler. To construct the coordinate system, the observer's inherent units at the present moment are first hypothetically duplicated to the whole space-time, resulting in a rigid and uniform Minkowskian coordinate system. Taking this coordinate system as the background to geometrize gravitation, the difference between the local inherent units of space-time and the observer's inherent units of space-time constitutes the curvature of space-time. Therefore, the coordinates in the space-time metric have clear physical meanings and they are directly related to the observer's physics.\\
3, According to Sections (5) and (6), above explanation of the nature of inertial forces directly proves that Einstein's strong equivalence principle is incorrect, and above success in generalizing the principle of dynamical relativity proves that the thorough general principle of relativity, at least for the time being, is insufficiently trustworthy until the basic requirement of causal symmetry is satisfied. As a result, real clocks that are free-falling in a gravitational field (including co-moving points in cosmology) are no longer synchronized(here by default after subtracting the observational time dilation effect in special relativity due to clocks' instantaneous velocities). Instead, there exists a discrepancy in their clock running rates,, which is dependent on the potential energy of the gravitational field in their respective locations. Because gravitational experiments in the solar system are remarkably successful, Einstein gravitational field equation is retained as the correct mathematical formula for the geometrization of gravitation. However, to satisfy the basic requirement of causal correspondence, the geometrization of gravitation must be based on an actual reference frame, and the energy-momentum tensor must also be accounted for relative to the actual reference frame. For instance, the geometrization in the Schwarzschild metric is based on a heliocentric reference frame. Therefore, the counting of the energy-momentum tensor should, in principle, not take into account the distribution of matter outside the solar system. This is precisely what the actual solution process follows by default.

Given that the observer is located at an infinite distance from the Sun, the solar system's geometrical description of gravitation is given by the Schwarzschild metric\cite{Chenchi}
\begin{equation}
ds^2=-(1-\frac{2GM}{r})dt^2+(1-\frac{2GM}{r})^{-1}dr^{2}+r^{2}(d\theta^{2}+sin^{2}\theta d\phi^{2}).
\end{equation}
Notably, now the time coordinate $t$ has a clear physical meaning. In Schwarzschild metric, the running rate of  the time coordinate $t$ is always defined by the running rate of the real clock (at the present moment) equipped by the observer at infinity and is duplicated into full space. Therefore, the clock corresponding to the time coordinate $t$ can be called the observational clock. Similarly, the basic unit of space coordinate $r$ is also directly linked to an observer at infinity. In the above expression, the inherent units of $dt,dr$ are defined by the observer and therefore can be understood as they are measured by the observational clock and observational ruler. The gravitational time dilation effect is demonstrated by the difference between the reading numbers of the local clock ($d\tau$) and the observational clock ($dt$) in the same time background segment ($\overline{dt}$): $d\tau=\sqrt{1-\frac{2GM}{r}}dt$. Obviously, on the surface of the sun, $\sqrt{1-\frac{2GM}{r}}<1$. Therefore, corresponding to the same time background segment ($\overline{dt}$), the reading number of the local clock on the surface of the sun ($d\tau$) is smaller than the reading number of the observational clock ($dt$). In other words, the clock on the surface of the sun runs slower than the clock of the observer at the infinite distance.

\section{New Proposal 2/2: Necessary supplements to the cosmological metric and qualitative questioning of the proposition of dark energy}
Hubble's law of cosmic expansion states that the speed at which galaxies move away from us is directly proportional to their distance. This relationship is better understood in the context of the absolute background of cosmic space. The adjusted fundamental physical picture of space-time unifies the absolute background and the relative inherent units as a pair of complementary opposites. The background of cosmic space is considered innately homogeneous and isotropic because, in itself, there is no any matter in it. In contrast, galaxies and matter move above the background of cosmic space. However, their position and speed relative to the background cosmic space cannot be directly determined, making all measurable speeds relative. The knowledge that any measured speed is necessarily relative must be maintained throughout the reasoning process. Thus, if the speeds of every two neighboring galaxies at an equal distance are the same, it is natural for two distant galaxies to satisfy the direct proportionality of their relative speed and relative distance. This is because only the space with an absolute background will not be compressed, and therefore has the above logical relationship to naturally explain Hubble's law.

It is known that the density of matter has changed significantly from the beginning of the universe to the present day. Therefore, the potential of the average gravitational field in the universe, although uniform in space, varies over time. Suppose that in a purely gravitational universe, the expansion process should slow down as the kinetic energy of the expanding universe decreases. The decrease of the kinetic energy is transformed into an increase in its potential energy of average gravitational field. Based on the gravitational time dilation effect which has been verified in the solar gravity test, there must be an evolution in the running rate of the local clock at every commoving point of the universe. Therefore, the time coordinates in the cosmological metric must strictly distinguish the running rates between the Earth observer's clock and the comoving local clock, as far as the long time span is concerned in cosmology\cite{Chenchi}. The relationship of the running rates between these two clocks can be described by $d\tau=b(t)dt$. Therefore, a complete and self-consistent cosmological metric should be represented as follows,
\begin{equation}
ds^2=-b^{2}(t)dt^2+a^2(t)[\frac{dr^{2}}{1-kr^{2}}+r^{2}d\theta^{2}+r^{2}sin^{2}\theta d\phi^{2}].
\end{equation}
Here the time coordinate $t$ also has a clear physical meaning. Although it does not directly correspond to a real clock, its clock running rate is duplicated from the clock running rate of the Earth observer's real clock at the present moment. And it serves as the sole benchmark for future assessment and comparison of the running rate with all other real clocks. Therefore in cosmology, the clock corresponding to the time coordinate $t$ that is always copied according to the running rate of the real clock equipped by the observer at the present moment is called the observational clock. It can be mathematically verified that the above form is actually the most general cosmological metric that satisfies the cosmological principle. By further incorporating Einstein's gravitational field equation, the fundamental dynamics equations of cosmology can be written as follows,
\begin{eqnarray}
\frac{\dot{a}^{2}}{a^{2}b^{2}}+\frac{k}{a^{2}}&=&\frac{8\pi G}{3}\rho\\
\frac{1}{b^{2}}(\frac{\ddot{a}}{a}-\frac{\dot{a}\dot{b}}{ab})&=&-\frac{4\pi G}{3}(\rho+3p).
\end{eqnarray}

Since the acceleration is not a scalar, the value of acceleration can vary significantly in different observers' reference frames. In the cosmic observational coordinate system defined by the Earth observer at the present moment, the mathematical expression describing the acceleration of cosmic expansion is no longer $\frac{d^{2}a}{d\tau^{2}}$, where $\tau$ is measured by the local clock at an early time point under investigation. Considering that the redshift values of all light signals in the universe are measured by the Earth observer's clock at the present moment, i.e., measured by the coordinate time $t$, the truly observed acceleration of cosmic expansion should be described by $\frac{d^{2}a}{dt^{2}}$. The relationship between these two accelerations satisfies.
\begin{eqnarray}
\frac{d^{2}a}{d\tau^{2}}&=&\frac{1}{b^{2}(t)}\frac{d^{2}a(t)}{dt^{2}}-\frac{1}{b^{3}(t)}\frac{da(t)}{dt}\frac{db(t)}{dt}  \cr
&=& \frac{a}{b^{2}}(\frac{\ddot{a}}{a}-\frac{\dot{a}\dot{b}}{ab}).
\end{eqnarray}

Now, we make a physical analysis referring to the physics demonstrated by the solar gravitational experiments and Schwarzschild metric. In order to qualitatively illustrate the physical effect resulting from the increase of gravitational potential energy over time, we can examine the evolutionary property of $b(t)$ by analogy with the gravitational time dilation effect in the Schwarzschild metric. In the Schwarzschild metric, the factor($\sqrt{1-\frac{2GM}{r}}$) of the running rate of the local clock increases with the distance $r$, which is equivalent to increase with the gravitational potential energy.

In cosmology, the potential energy of average gravitational field in the early universe was low, whereas at the present moment the potential energy of average gravitational field in the universe is high. Therefore, during the expansion of the universe, with the increase of gravitational potential energy, the time factor $b(t)$ will increase with time, i.e., $\frac{\partial b(t)}{\partial t}>0$ . On the other hand, it is easy to see that during the expansion of the universe, $\frac{\partial a(t)}{\partial t}>0$, and thus, even if we obtain $\frac{\partial^{2} a(t)}{\partial t^{2}}>0$ based on existing observational data, according to the equation (20), we could still obtain a negative $\frac{\partial^{2} a(\tau)}{\partial \tau^{2}}$, and consequently, according to the equation (19), we could still have a positive $\rho+3p$. Therefore, it has been proven that the introduction of a physically complete cosmological metric of the Earth observer at the present moment is meaningful for correctly determining the acceleration of cosmic expansion\cite{Machprinciple}.

Furthermore, we can intuitively and qualitatively analyze the physical impact of gravitational redshift effects on the judgment of the acceleration of cosmic expansion. Analogous analyses are presented in Table 1 below for reference,
\begin{table}[h]
\begin{centering}
\caption{Qualitative analysis referring to the physics in solar gravitational experiments}\label{tab1}
\scalebox{0.9}{
\begin{tabular}{|p{9pc}|p{15pc}|p{4.5pc}|p{4.5pc}|}
\hline
Schwarzschild metric & $b(r)=(1-\frac{2GM}{r})^{\frac{1}{2}}$, the gravitational time dilation effect is verified & $b(r)>0$ & $\frac{db(r)}{dr}>0$ \\\hline
Cosmological metric & $b(t)=(1-\frac{\tilde{k}}{a(t)})^{\frac{1}{2}}$, as a speculative analogy & $b(t)>0$ & $\frac{db(t)}{dt}>0$ \\\hline
Two accelerations & even $\frac{d^{2}a(t)}{dt^{2}}>0$, but $\frac{d^{2}a(\tau)}{d\tau^{2}}<0$ is still possible since $d\tau=b(t)dt$ & $a(t)>0$ & $\frac{da(t)}{dt}>0$ \\\hline
\end{tabular}}
\end{centering}
\end{table}

By assuming specific values for the time dilation factor, qualitative analysis of the physical impact on the judgement of cosmic expansion acceleration may also be simulated, as illustrated in Table 2(The numerical values here are merely simulated),
\begin{table}[h]
\caption{Qualitative analysis by numerical simulation}\label{tab2}
{\centering
\scalebox{0.9}{
\begin{tabular}{|p{10pc}|p{7pc}|p{8pc}|p{11pc}|}
\hline
$t$: the coordinate time;      $\tau$: the local time & At an early moment: $1s=\frac{1}{2}s'$ & At the present moment:   $1s=1s'$ &Trend \\\hline
Coordinate expansion speed: &If $v=1m/s$ & If $v=1.5m/s$ & Apparently accelerating  \\\hline
Local expansion speed:  & Then $v=2m/s'$ & Then $v=1.5m/s'$ & Essentially decelerating  \\\hline
\end{tabular}
}}
\end{table}

\section{Conclusion}
This article discusses the physical impacts resulting from the revelation of the nature of inertial forces. The nature of the inertial force has been revealed through the successful generalization of particle dynamics to the translational reference frame(with non-rotating coordinate axes and arbitrary accelerated reference origin). The generalized dynamics equation (10) remains invariant under arbitrary transformation between translational reference frames. This introduces a moderate principle of relativity. In comparison, the various principles of dynamical relativity are summarised in Table 3,

\begin{table}[h]
\caption{Summary of principles of dynamical relativity}\label{tab3}
{\centering
\scalebox{0.85}{
\begin{tabular}{|p{9pc}|p{12pc}|p{17pc}|}
\hline
Galilean Principle of Relativity & the dynamics law remains the same form in any inertial reference frame & Physical basis: Newton's absolute view of space-time(Galilean transformations of velocity)    \\\hline
Special Principle of Relativity & Physical laws are formally invariant in any inertial reference frame & Physical Basis: Relativistic view of space-time(principle of invariant speed of light)   \\\hline
General Principle of Relativity & Physical laws are formally invariant in any frame of reference & Physical Basis: Einstein's Strong Equivalence Principle(gravitational and inertial forces are physically equivalent)    \\\hline
Moderate Principle of Relativity & The dynamics law remains the same form in any translational reference frame& Physical Basis: The principle of causal symmetry for reference object(s)    \\\hline
\end{tabular}
}}
\end{table}

According to the formula (10), which is strictly derived from Newtonian dynamics, the inertial force in a translational reference frame is essentially the real force acting on the reference object upon which the reference frame is defined. Therefore, from a physical perspective, inertial forces may not only be gravitational, but also be various other common forces. Therefore, the inertial force is actually not equivalent to the gravitational force. More importantly, inertial forces do not actually act on the object under study, but rather on the reference object. Currently, only the gravitational force has been found to exhibit the time dilation effect. This fundamental explanation of inertial forces contradicts Einstein's strong equivalence principle. It is well known that in the general theory of relativity, Einstein has eliminated the direct dependence on inertial reference frames. Now, however, the physical clue that supported this idea lost its crucial logical support.

The understanding of inertial forces also remains valid when the principle of dynamical relativity is desired to be generalized to rotating reference frames. It is important to ensure that the forces acting on the reference objects and the acceleration of the corresponding reference frame satisfy causal symmetry and consistency. Specifically based on the analysis of degrees of freedom, the construction of the general principle of relativity requires causal symmetry between the kinematics and all the forces acting on at least four reference objects. However, formulating a unified formula incorporating totally five different objects' dynamics is too difficult in mathematics (namely 1 object under study + 4 reference objects). Therefore, the general principle of relativity in particle dynamics has never been actually realized within the current theoretical framework. Nevertheless, the only correct way to generalize the principle of dynamical relativity can be demonstrated as follows in Table 4,

\begin{table}[h]
\caption{Particle dynamics with causal symmetry}\label{tab4}%
\begin{centering}
\scalebox{0.85}{
\begin{tabular}{@{}llll@{}}
\toprule
the empirical law & applicable reference frames& physics introduced& presentations\\
\midrule
& Inertial reference frames & 0 reference particle  &${\textbf {\emph F}}\vert_{p}=m_{p}{\textbf {\emph a}}\vert_{p-\Sigma} $  \\
$\Delta{\textbf {\emph F}}\cong m\Delta{\textbf {\emph a}}$  & Translational reference frames & 1 reference particle & $\frac{{\textbf {\emph F}}\vert_{p}}{m_{p}}-\frac{{\textbf {\emph F}}\vert_{o}}{m_{o}}={\textbf {\emph a}}\vert_{p-O}$   \\
& Arbitrary reference frames  & 4 reference particles & unrealizable!?  \\
\bottomrule
\end{tabular}
}
\end{centering}
\end{table}

There is a necessity for generalizing the principle of relativity from inertial reference frames to translational reference frames, and this necessity might be illustrated by the following analogy presented in Table 5,
\begin{table}[h]
\caption{Necessity for correctly distinguishing real forces}\label{tab5}
\begin{centering}
\scalebox{0.8}{
\begin{tabular}{|c|p{13pc}|p{17pc}|}
\hline
Kinematics & Geocentric reference frame $\Rightarrow$ Heliocentric reference frame & Correctly distinguish between the motions of the celestial bodies or the Earth's frame of reference itself   \\\hline
Dynamics & Inertial reference frame $\Rightarrow$ Translational reference frame & Correctly distinguish between the forces acting on the investigated object or the reference object    \\\hline
\end{tabular}
}
\end{centering}
\end{table}

Historically, the revolution from the geocentric to the heliocentric theory played a crucial role in clarifying the actual motion of celestial bodies. This, in turn, led to the discovery of Kepler's three laws of motion and eventually Newton's law of universal gravitation. Drawing an analogy, it is similarly important to clarify real forces acting on objects, as this may influence the understanding of the forces acting on them.

Understanding the correct causality behind the acceleration in particle dynamics has a significant impact on the dynamics of the universe. In particle dynamics, acceleration is directly related to force, thus revealing the nature of inertial forces may ultimately lead to a correct assessment of the acceleration of the expansion of the universe.

In fact, the thorough re-derivation of the generalized dynamics equation in Section 4 has reflected the true process how we discover the nature of inertial forces. The re-derivation is based on the objective position of each particle in the background of cosmic space at any given moment, which is consistent with the concept of ``events" in special theory of relativity. However, there is an underlying logic hat is equally fundamental to special relativity but has been overlooked. This ``objectivity" has substantially implied that the background of cosmic space is absolute. The concept of space-time should be further subdivided into the background of space-time and the inherent units of space-time. The background of cosmic space is a void that remains after the removal of all matter and energy from the universe. Specifically, there is no any form of interaction that can act upon the background of cosmic space, so it must be absolute. The inherent units of space-time, which is defined by the inherent physical phenomena of specific matter, can be influenced by interactions and can therefore be relative. In addition, the ``absoluteness" of the background of cosmic space can also be tested by a rigorous and high-precision experiment on the generalized dynamical equation (10). If the equation is found to be biased in high-precision experiments, then possible corrections should come from three sources: 1, the presence of unknown new interactions or forces that are not taken into account; 2, the potential incorrectness of the microscopic causality between force and acceleration; and 3, the initially determined ``background" of cosmic space is not absolute, but may actually have its own dynamics.

Regarding general relativity, the new interpretation of the nature of inertial forces presented in this article completely deviates from Einstein's strong equivalence principle. The proposed approach to generalize the principle of relativity also illustrates the infeasibility of the general principle of relativity under the current mathematics. Therefore, it is proposed to abandon the strong equivalence principle and the general principle of relativity. Furthermore, the equivalence of freely falling clocks in different gravitational fields is no longer valid, given that the gravitational force is no longer equivalent to the inertial force. This necessitates a modification of the physical picture of how the gravitational force affects the geometry of space-time. In a more conservative new physical picture, both time and space coordinates have clear physical meanings, and their basic units are defined according to the running rate of the real clock and the background length of the real ruler, respectively, equipped by the observer at the present moment. Furthermore, as the observer at the present moment is always unique, the coordinate system, which serves as the reference basis for gravitational geometrization, is originally homogeneous and isotropic, and is curved by the influence of gravitational fields. Besides, it is recommended to retain Einstein's gravitational field equation since it is primarily based on Einstein's weak equivalence principle and has been extensively verified by solar gravitational experiments.

The gravitational redshift effect in the solar system can be naturally explained by the modified physical picture of space-time. And according to the physical consistency, the cosmological metric in observational theory should be modified to a more general formula (17).
As the universe expands, the decrease in its kinetic energy results in an increase in the potential energy of the average gravitational field in the universe. This leads to an acceleration of the running rate of the local clock in the universe in the longitudinal direction of time. The factor $b(t)$ introduced in the generalized cosmological metric (17) just describes this physics. Therefore, making a coordinate transformation $d\tau=b(t)dt$ would be physically meaningless. Here the time coordinate $t$ based on the running rate of the Earth observer's real clock at the present moment is observationally meaningful. Since all of our observations of light signals from the early universe were made by Earth observers at the present moment, the redshift values of these light signals were determined by comparing them to the radiation of similar atoms occurring on Earth at the present moment. As a result, our observed acceleration of the expansion of the Universe is actually the coordinate acceleration, which is measured by the Earth observer's clock running rate at the present moment. It is crucial to emphasise that this characterization is applicable to all arbitrarily early cosmic accelerations as long as they are observed at the present moment. Such an observed acceleration must be distinguished in cosmology from the local acceleration, since only the local acceleration of the expansion is measured by the running rate of the local clock at the same moment in the early universe (i.e., when the light signal was emitted). According to equation (19), it is only the local acceleration that really corresponds to the criterion for judging the existence of dark energy. Ultimately, the qualitative analysis suggests that even if current cosmological observations are correct, the crisis of the accelerating expansion of the universe can be effectively mitigated. At the very least, the dark energy hypothesis should be re-examined.

\section*{Acknowledgements}

Thanks to the support of Researcher. Shen Yougen, Researcher Zhang Xinmin, and Prof. Li Kang. Without delving into the cosmological metric and exploring its physical origins since the end of 2004,  the author would not have been able to trace back to the genesis of this work. This study has been funded by the Natural Science Foundation of Zhejiang Province (No. Y6110778). Additionally, this research has received partial support from the Project of Knowledge Innovation Program (PKIP) of the Chinese Academy of Sciences (No. KJCX2.YW.W10).

\end{document}